\documentclass[a4paper,aps,preprintnumbers,amsmath,amssymb,prl]{revtex4}
\usepackage{mathrsfs}
\usepackage{amsmath}
\usepackage[dvips]{graphicx}
\usepackage{epsfig}
\usepackage[dvips]{graphics,graphicx}
\begin{document}

\title{Theoretical Study of Oligophenyl-based Double Barrier Molecular Device}

\author{Aranya B Bhattacherjee$^{1}$ and Suman Dudeja$^{2}$ }

\address{$^{1}$Department of Physics, ARSD College, University of Delhi (South Campus), New Delhi-110021, India} \address{$^{2}$Department of Chemistry, ARSD College, University of Delhi (South Campus), New Delhi-110021, India}

\begin{abstract}
We report an investigation of electron conduction in oligophenyl based double barrier molecular device. We have carried out analytical calculations and numerical simulations on isolated molecules, consisting of aromatic $\pi$ conjugated system made up of three phenyl rings separated by insulator groups $-CH_{2}-$, $-SiH_{2}-$, $-GeH_{2}-$ and $-SnH_{2}-$. We show analytically as well as numerically that when the two insulator groups are different an asymmetric electron transport in the presence of external electric field is possible,thus resembling a diode which allows one way electron transport.
\end{abstract}

\maketitle

\section{Introduction}

Recent progress in experimental and theoretical techniques have triggered enormous possibilities of using organic molecules as a fundamental complement to silicon in semiconductors \cite{bumm, jones, reed}. This has opened up possibilities of fabricating logical circuits at the single molecule level \cite{wu}. Such molecular state electronics requires the precise knowledge of electron transport in organic molecules The finite size of atoms, molecules and clusters offers an advantage in designing tailor-made materials to perform desired functions by exploiting the quantum mechanical phenomena of electrons in a confined space.
Besides such exciting developments on the design and fabrication of molecular devices, experiments have been performed on the electrical properties of single molecules \cite{kerg}, and highly sophisticated simulation tools for charge transport through molecules have been developed \cite{brand, ventra, damle, taylor}. Recently, using a single benzene molecule, the first molecular transistor was developed \cite{song}.
Several molecular device concepts are based on the existence of potential barriers within molecules, which prevent the free flow of electrons and thus create a tunable system \cite{seminario1, seminario2, karzazi, majumdar, girlanda}. Aviram and Ratner \cite{aviram} proposed such a device in 1974 based on an organic molecule made of $\pi-$ donor and $\pi-$ acceptor moieties linked by a saturated spacer and sandwiched between two metallic electrodes. This concept has recently been verified experimentally \cite{metzger}. The presence of potential barriers is essential also in devices which are closer to traditional electronic electronic components.

A diode is a two-terminal switch consisting of source and drain that can turn a current on and off depending on the direction of flow. In the case of molecular diodes, switching the current 'on or off' requires tuning the molecular discrete energy levels. In an experiment in 1999 \cite{chen}, it was shown that a single molecule of $2'-amino-4-ethynylphenyl-4'-ethynylphenyl-5' nitro-1-benzene thiol$ could behave like a resonant tunneling diode. The resonance effect of tunneling the electrons was observed at $2.1V$ bias with a peak current of $1nA$. The molecule consists of three benzene rings connected to each other by the acetylene group. Two $H$ atoms of the central ring are replaced by the $NO_{2}$ and $NH_{2}$ group on opposite sides. This asymmetrical configuration enables the molecule susceptible to change its configuration under perturbation.

In the present work we are interested in a quantitative investigation of molecular structures containing aliphatic chains acting as barriers inserted between  aromatic $\pi-$ conjugated rings, focusing on systems made up of three phenyl rings separated by insulator groups $-CH_{2}-$, $-SiH_{2}-$ ,$-GeH_{2}-$ and $-SnH_{2}-$. We show analytically as well as numerically that when the two insulator groups are different an asymmetric electron transport in the presence of external electric field is possible,thus resembling a diode which allows one way electron transport. To test the barrier properties we have performed analytical quantum mechanical calculations and also performed numerical simulations. We have based our analysis on the analytical results obtained for the occupation probability of the energy levels, the delocalization of the Highest Occupied Molecular Orbital (HOMO) and the HOMO-LUMO energy gap (HLG). The HLG is a critical parameter for the molecular admittance because it is a measure of the hardness of the electron density \cite{seminario2}. The larger the HLG, the more stable the molecule, and therefore the harder it is to rearrange its electron density under the presence of an external electron. Thus the molecule presents lesser admittance to the incoming electron.

\section{The Model and Analytical Results}

\begin{figure}[t]
\hspace{-0.0cm}
\includegraphics [scale=1.0]{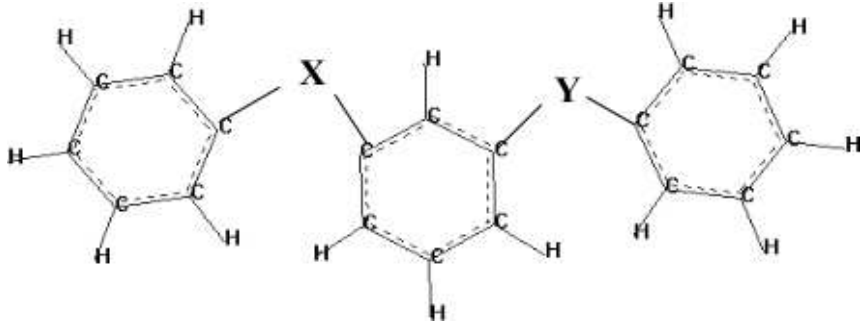}
\caption{ Schematic representation of the system we are investigating. The $X$ and $Y$ are any one of the insulating groups $-CH_{2}-$, $-SiH_{2}-$, $-GeH_{2}-$ and $-SnH_{2}-$. If $X \neq Y$ then an asymmetry is generated in the electron conduction process when an electric field is applied.}
\label{1}
\end{figure}

The source of conductivity for a polyphenylene wire is a set of delocalized $\pi-$ type molecular orbitals above and below the plane of a single benzene ring. In this section, we present a quantum mechanical analysis of electron density behaviour of the polyphenylene wire in which insulator groups $-CH_{2}-$, $-SiH_{2}-$, $-GeH_{2}-$ and $-SnH_{2}-$ are inserted between two aromatic rings and an external electric field is applied. The system considered is shown in Figure 1, where $X$ and $Y$ are the insulator groups $-CH_{2}-$, $-SiH_{2}-$, $-GeH_{2}-$ and $-SnH_{2}-$ . If $X \neq Y$ then an asymmetry is generated in the electron conduction process when an electric field is applied. We now present here a simple three state quantum mechanical model which provides an intuitive understanding of the role played by the aliphatic fragments in the triphenyl system.

The Hamiltonian for a single electron in a system with three levels $E_{1}$, $E_{2}$ and $E_{3}$, coupled through an external perturbation, can be written, in second quantized formalism, as

\begin{equation}
H=\sum_{i=1,2,3}n_{i} E_{i}+\sum_{i=1,2,3} q n_{i} V_{i}+t_{1}(b_{1}^{\dagger}b_{2}+b_{2}^{\dagger}b_{1})+t_{2}(b_{2}^{\dagger}b_{3}+ b_{3}^{\dagger}b_{2}),
\end{equation}

where $n_{i},i=1,2,3$ are the occupation numbers of the three levels accessible to the electron, $b_{i}^{\dagger}$ and $b_{i}$ are the creation and annihilation operators for an electron in the $i^{th}$ level, $t_{1}$ and $t_{2}$ represents the interlevel coupling or the tunneling matrix element. The second term takes into account the perturbing potential $V$ through its value $V_{i}$ in the location corresponding to the $i^{th}$ level. If the Hamiltonian is expanded in the following simple basis

\begin{equation}
|1>, n_{1}=1, n_{2}=0, n_{3}=0,
\end{equation}

\begin{equation}
|2>, n_{1}=0, n_{2}=1, n_{3}=0,
\end{equation}

\begin{equation}
|3>, n_{1}=0, n_{2}=0, n_{3}=1.
\end{equation}

and written in matrix form, it reads, in atomic units, as

\begin{equation}
H = \left ( \begin{array}{ccc}
 {E_{1}+q V_{1}} &  {t_{1}}& {0} \\ {t_{1}} & {E_{2}+q V_{2}} & {t_{2}} \\ {0} & {t_{2}} & {E_{3}+q V_{3}}
\end{array}  \right )
\end{equation}

We now define the energies $\epsilon_{1}=(E_{2}+q V_{2})-(E_{1}+q V_{1})$ and $\epsilon_{2}=(E_{3}+q V_{3})-(E_{1}+q V_{1})$ and obtain the three eigenvalues of the above Hamiltonian as the roots of the following cubic equation

\begin{equation}
\lambda^{3}-\lambda^{2}(\epsilon_{1}+\epsilon_{2})+\lambda(\epsilon_{1} \epsilon_{2}-t_{1}^{2}-t_{2}^{2})+\epsilon_{2}t_{1}^{2}=0
\end{equation}

Now for the simple case $\epsilon_{2}=0$ and $\epsilon_{1}=\epsilon$, we obtain two eigenvalues as

\begin{equation}
\lambda_{1}=\frac{\epsilon+\sqrt{\epsilon^{2}+4 (t_{1}^{2}+t_{2}^{2})}}{2}
\end{equation}

\begin{equation}
\lambda_{2}=\frac{\epsilon-\sqrt{\epsilon^{2}+4 (t_{1}^{2}+t_{2}^{2})}}{2}
\end{equation}

\begin{figure}[t]
\hspace{-0.0cm}
\includegraphics [scale=1.0]{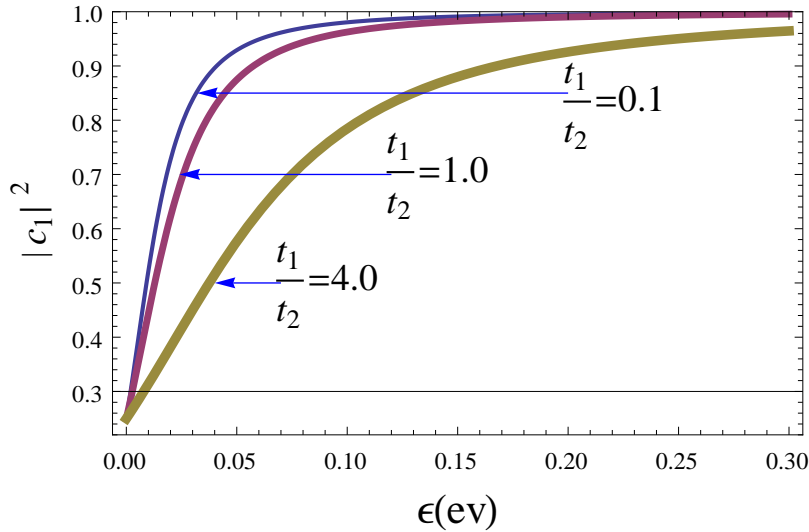}
\caption{Occupancy of the first level of the analytically solvable model as a function of the perturbation $\epsilon$, for three different values of the ration $t_{1}/t_{2}$. }
\label{2}
\end{figure}

In Figure 2 we plot the occupancy of the first level $|c_{1}|^{2}$ as a function of $\epsilon$ for three values of the ration $t_{1}/t_{2}$. Here $c_{1}$ is the first coefficient of the eigenvector of the ground state . The analytical expression for $c_{1}$ is

\begin{equation}
c_{1}=\frac{\lambda_{1}^{2}}{(\lambda_{1}^{2}+t_{1}^{2}+t_{2}^{2})}
\end{equation}

The parameters $t_{1}$ and $t_{2}$ describes the effective interaction between the orbitals of the three aromatic rings and therefore will be smaller when an aliphatic group breaks the $\pi$-conjugation and spaces the rings. An asymmetry is produced when $t_{1} \neq t_{2}$ . In Figure 2, when $t_{1}/t_{2}=4.0$, the single electron will find it easier to tunnel from the first ring to the middle ring through the insulator $X$ as compared to the path starting from the middle ring to the third ring via insulator $Y$. Since $t_{1}>t_{2}$, the internal barrier between the first ring and the middle ring is small and the aromatic fragments are closer together and strongly interacting. On the other hand when $t_{1}/t_{2}=0.1$ exactly opposite is true and hence the electron density will tend to localize on the first ring as evident from Figure 2 when $|c_{1}|^{2}\rightarrow 1$ at a smaller value of $\epsilon$ compared to the cases $t_{1}/t_{2}=1.0$ and $t_{1}/t_{2}=4.0$. This asymmetry that is introduced when $t_{1} \neq t_{2}$ can be utilized to design a tunable molecular tunneling diode. The curve in Figure 2 for the highest value of $t_{1}/t_{2}$ exhibits the desired unidirectional (left to right) flow of electron current if initially the electron is localized on the left ring.

\section{Numerical Results}

\begin{figure}[t]
\hspace{-0.0cm}
\begin{tabular}{c}
\includegraphics [scale=0.8]{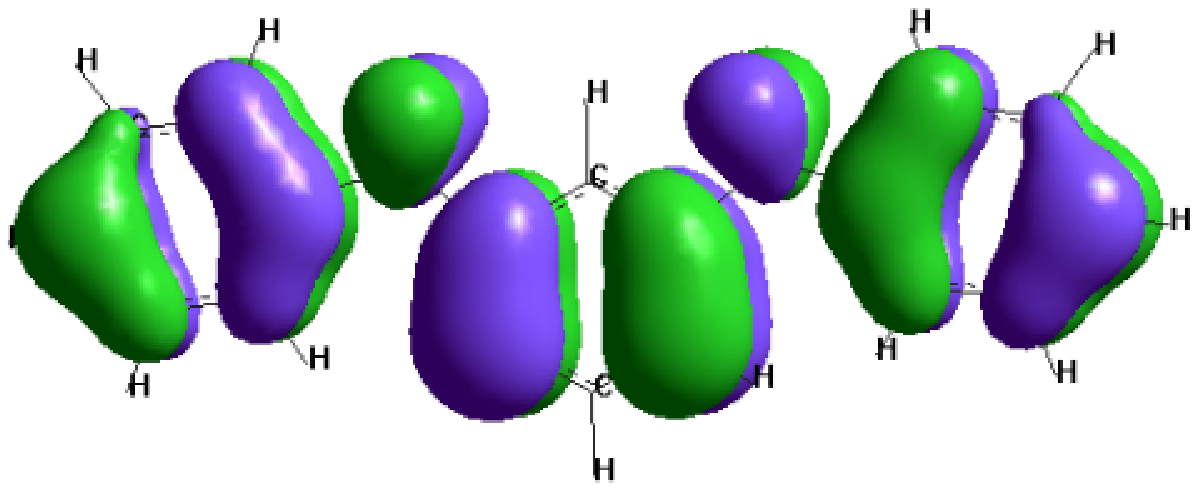}\\
\includegraphics [scale=0.8] {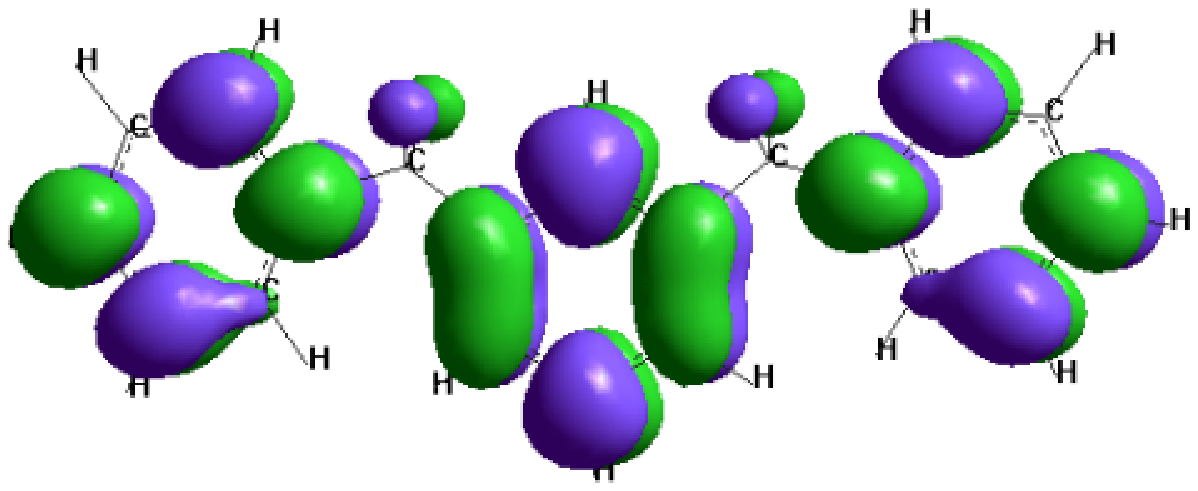}
\end{tabular}
\caption{Optimized structure of the $TP-CH_{2}-$ system with the HOMO distribution(top) and the extend of the LUMO orbitals (bottom). Symmetrical distribution of HOMO and LUMO is visible.}
\label{3}
\end{figure}

\begin{figure}[t]
\hspace{-0.0cm}
\includegraphics [scale=0.8]{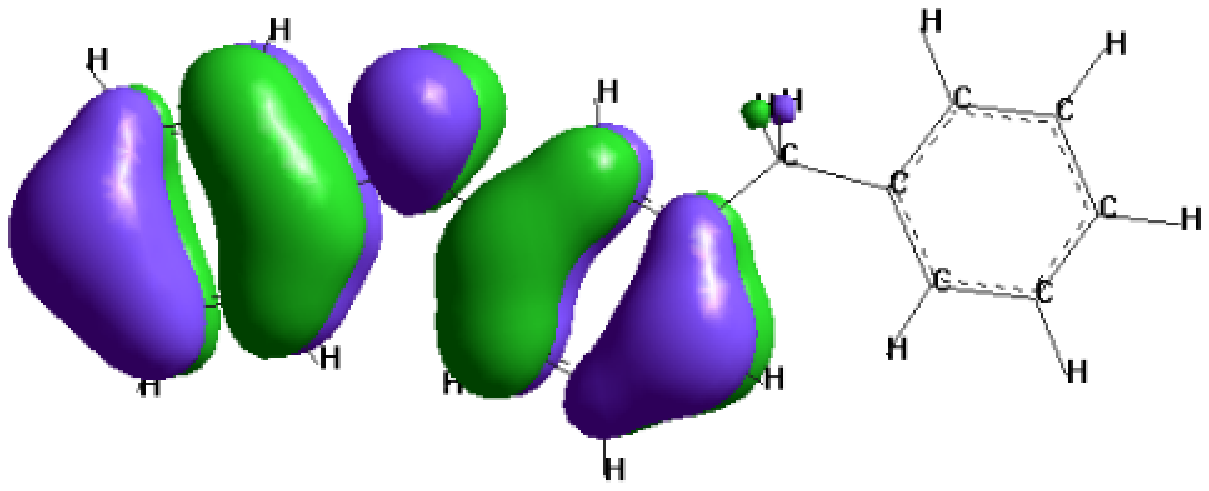}\\
\includegraphics [scale=0.8] {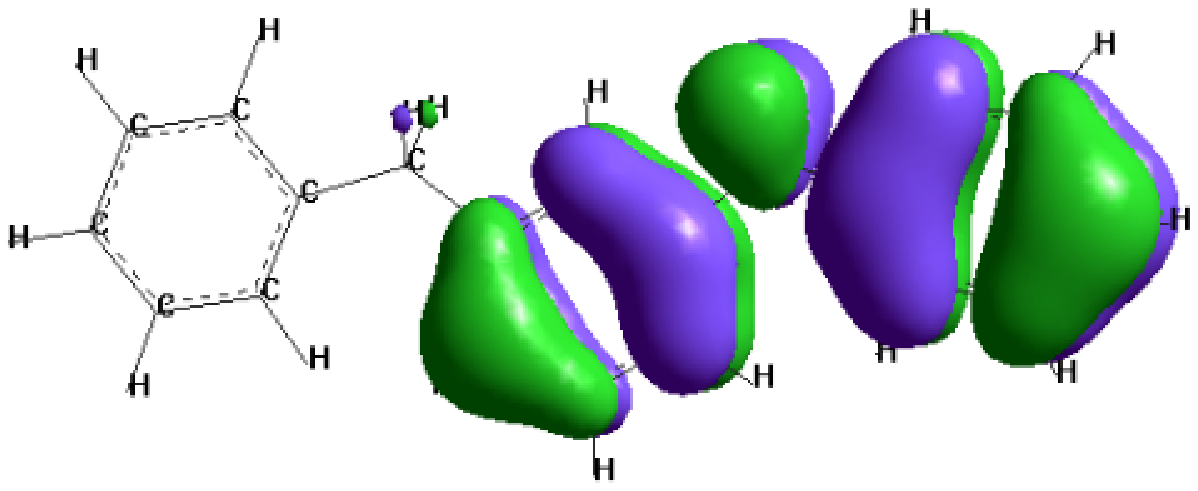}
\caption{Shape of the HOMO orbitals of the $TP-CH_{2}-$ system in the presence of electric field $E=\pm 0.001$ atomic units. $1 $ atomic unit $=5.142 \times 10^{11} V/m$.  }
\label{4}
\end{figure}

All calculations were performed using Hyperchem $7.0$ software. To obtain good starting geometries, the initial configurations were optimized at the Hartree-Fock (HF) level. The density functional theory (DFT) method was used on the optimized geometries obtained from HF calculations. We have used the $6-31G^{*}$ basis and hybrid functional $B3PW91$, which combines the Becke exchange (B) functional, and the Perdew-Wang-91 correlation functional, both of which are nonlocal generalized gradient-approximated functionals. Changes in the molecular orbitals are excellent indicators of the molecular electron transport. It is important to understand the electronic molecular transport under the action of an external electric field along the axis of the molecule. In this context the HLG energy and the spatial extent of the HOMO are good candidates. We need to quantify two effects regarding the transport of electrons in molecules to investigate their electrical properties. First, the HOMO effect needs to be quantified because the charge transfer from one end of the molecule to the other end becomes harder to achieve when the HOMO is delocalized over the entire molecule. The second effect is due to a directly quantifiable property, the HLG. When this gap decreases, the molecular admittance increases. First, we analyze the case when $X=Y=-CH_{2}-$. Figure 3 shows the extend of the HOMO and LUMO orbitals for the case $X=Y=-CH_{2}-$.  Clearly we notice the symmetrical nature of the extend of HOMO and LUMO across the molecule due to the symmetric nature of electron withdrawing and electron donating groups. This is the case of $t_{1}=t_{2}$. The geometry optimization was performed initially by the HF and then by the DFT method. Interestingly the two methods yield different HLG energy. Table 1 shows the comparison of the electronic energy levels calculated using HF and DFT methods for the molecule triphenyl system with two $-CH_{2}-$ spacers($TP-CH_{2}-$).

\bigskip
\bigskip

\begin{tabular}{ccccc}
Table 1: Comparison &
of the electronic energy&
levels of $TP-CH_{2}-$ &
calculated using &
HF and DFT methods. \\

\hline
\multicolumn{1}{c}{Method} &
\multicolumn{1}{c}{$E_{HOMO}$ (eV)} &
\multicolumn{1}{c}{$E_{LUMO}$ (eV)} &
\multicolumn{1}{c}{$E_{LUMO+1}$ (eV)}&
\multicolumn{1}{c}{HLG(eV)} \\
\hline
DFT & $-4.73$ & $2.34$ & $2.39$ & $7.07$ \\
HF & $-9.12$ & $0.42$ & $0.43$ & $9.54$
\end{tabular}

\bigskip
\bigskip

\begin{figure}[t]
\hspace{-0.0cm}
\includegraphics [scale=0.8]{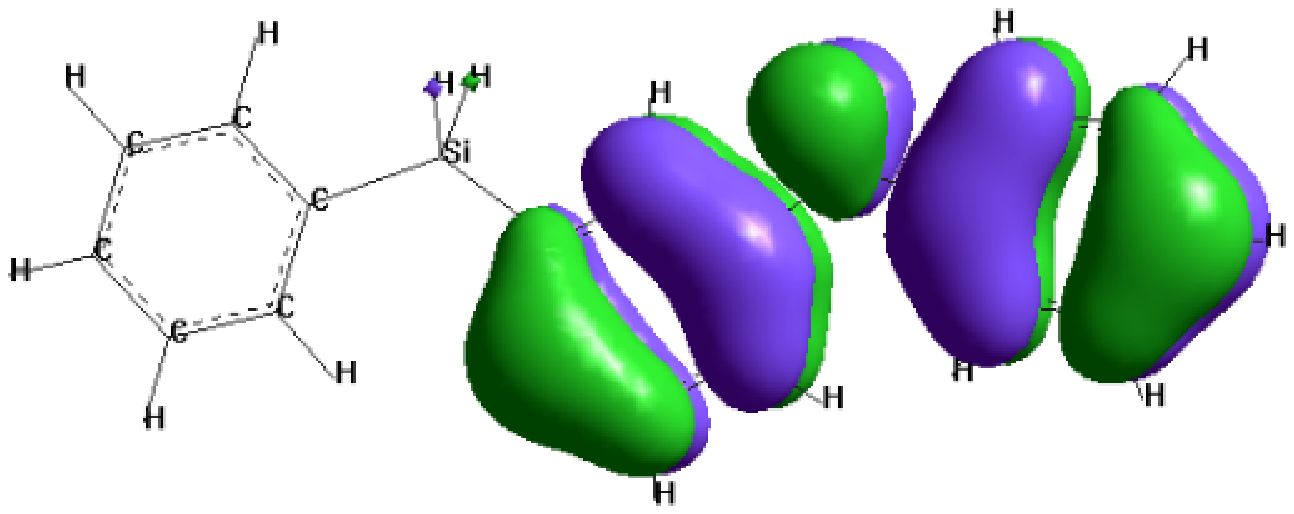}\\
\includegraphics [scale=0.8] {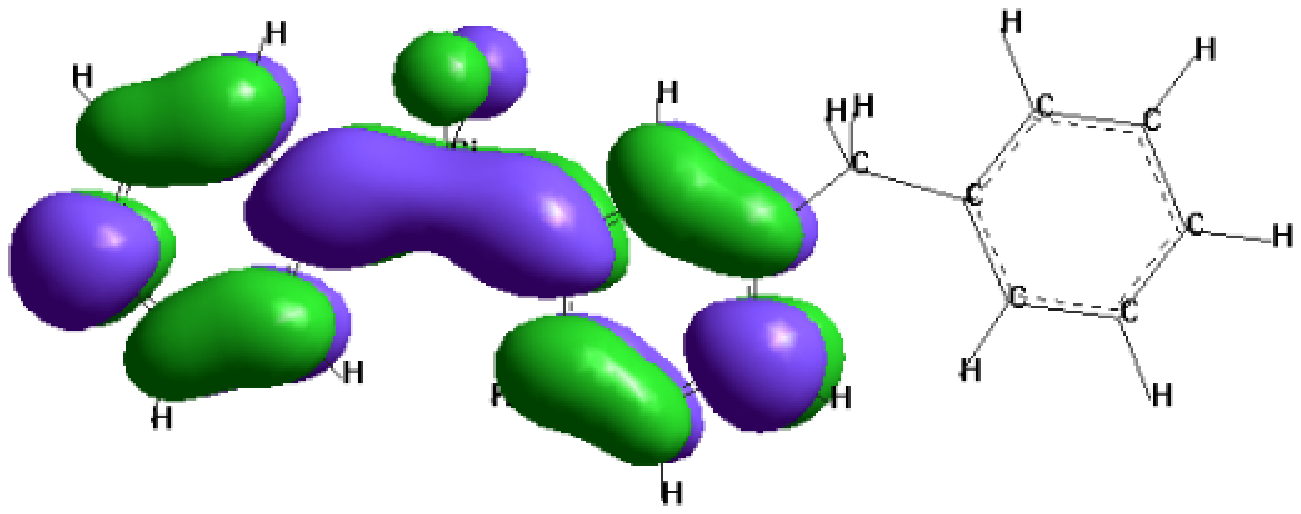}
\caption{Optimized structure of the $TP-SiH_{2}-$ system with the HOMO distribution (top) and the extend of the LUMO orbitals (bottom). Asymmetrical nature of the LUMO and the HOMO is visible. The distance between the two phenyl rings via the $-CH_{2}-$ spacer is $2.669 {\AA}$ and via the $-SiH_{2}-$ spacer is $3.136 {\AA}$. This gives rise to the asymmetrical tunneling matrix element.}
\label{5}
\end{figure}

\begin{figure}[t]
\includegraphics [scale=0.8]{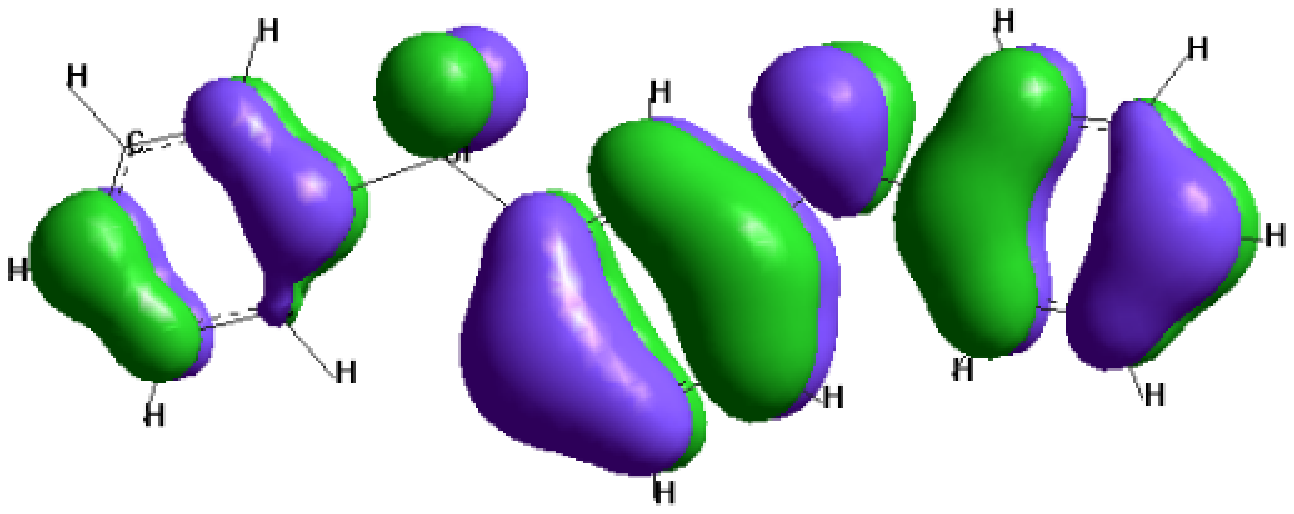}\\
\includegraphics [scale=0.8] {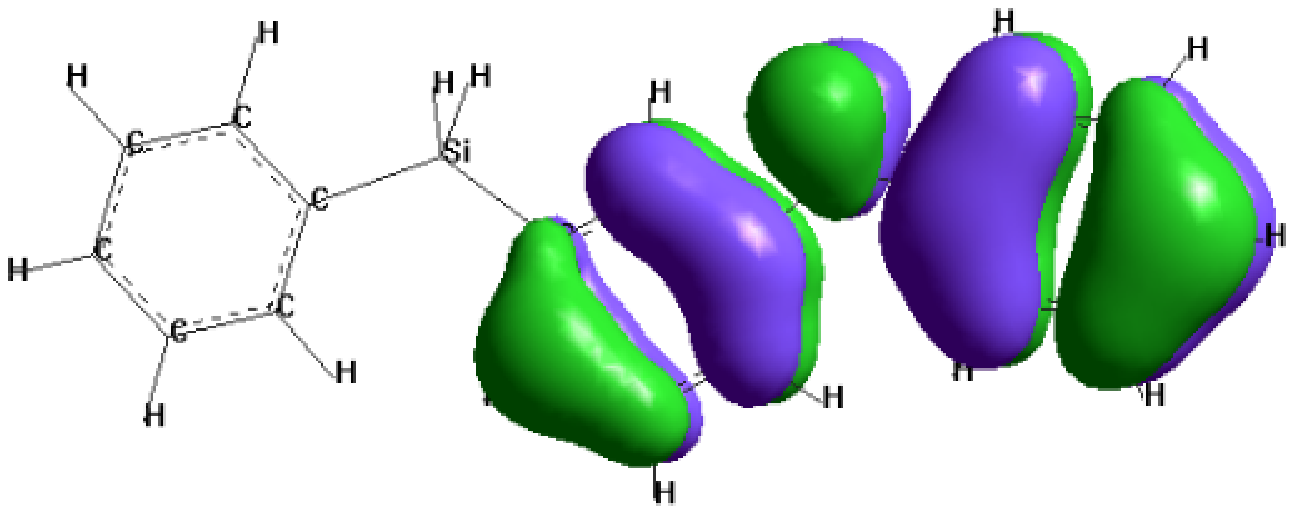}
\caption{Shape of the HOMO orbitals of the $TP-SiH_{2}-$ system in the presence of electric field $E=\pm 0.001$ atomic unit. $1$ atomic unit $=5.142 \times 10^{11} V/m$. The top figure is for $E=0.001 $ atomic unit and the bottom figure is for $E=-0.001$ atomic unit.  }
\label{6}
\end{figure}

A significant difference in the HLG is noticed from Table 1. The reason for such a difference in the results obtained from HF and DFT method is due to the absence of electron correlations in the HF method. Electron correlation effects are expected to play important role in this molecule. Any incoming electron from the metallic electrodes connected to the one end of the molecule is expected to propagate over the $LUMO+1$ under a suitable bias voltage required to transfer one electron through this molecule. We assume in the first approximation that the Fermi level of the metal electrode lies in the middle of the HOMO and LUMO energy levels of the molecule. To determine the bias voltage required ($E_{v}$)to transfer one electron through this molecule is estimated as \cite{majumdar}

\begin{equation}\ref{ev}
E_{v}=\frac{HLG}{2}-(E_{LUMO}-E_{LUMO+1})
\end{equation}

Here $HLG$ represents the energy gap between HOMO and LUMO energy levels. $E_{LUMO}$ and $E_{LUMO+1}$ represent the energy for the $LUMO$ orbital and $LUMO+1$ orbital, respectively. The $E_{v}$ for the molecule $TP-CH_{2}-$ from the above Eqn.(10) is estimated to be $3.59 eV$.

Figure 4 shows the spatial orientation diagram for the HOMO energy levels of the molecule $TP-SiH_{2}-$ (triphenyl system with one $-CH_{2}-$ and one $-SiH_{2}-$ spacer) in the presence of electric field $E=\pm 0.001au$. Due to the symmetry of the molecule, the HOMO energy levels are symmetrically displaced around the central phenyl ring. This is a consequence of the fact that $t_{1}=t_{2}$.

\begin{figure}[t]
\hspace{-0.0cm}
\includegraphics [scale=1.0]{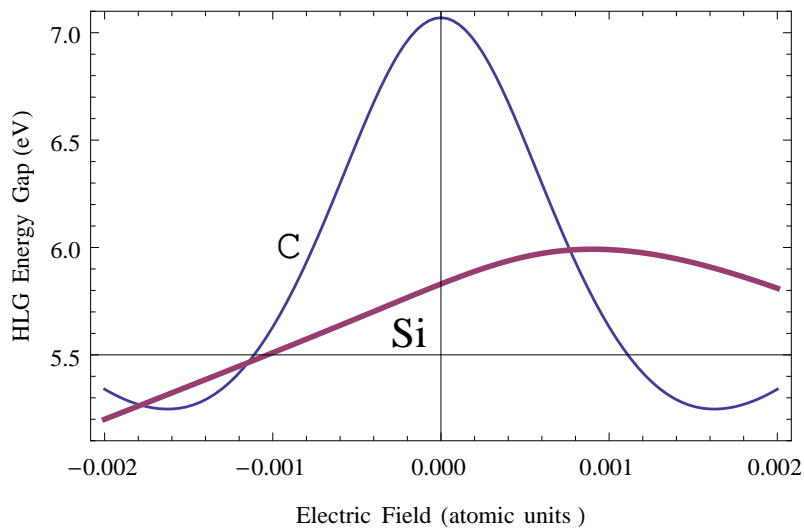}
\caption{Plot of the energy gap $HLG$ versus electric field. For the molecule $TP-CH_{2}-$, the plot is symmetric indicating the absence of any MRTD behaviour. On the other hand, for the second molecule $TP-SiH_{2}-$, the electron experiences an energy barrier on the positive electric field side. }
\label{7}
\end{figure}

When the insulator spacer is not identical ($X \neq Y$), the situation varies radically. Figure 5, displays the double barrier molecular RTD for the case when $X=-SiH_{2}-$ and $Y=-CH_{2}-$ giving rise to $t_{1}\neq t_{2}$ and $t_{1}<t_{2}$ (The distance between the two phenyl rings via the $-CH_{2}-$ spacer is $2.669 {\AA}$ and via the $-SiH_{2}-$ spacer is $3.136 {\AA}$. This gives rise to the asymmetrical tunneling matrix element. A larger distance indicates a smaller tunneling matrix element).  This asymmetric charge distribution gives rise to asymmetric spatial orientation of the HOMO and LUMO energy levels. This is exactly what one requires to design a double barrier molecular resonant tunneling diode. Figure 6 shows the spatial orientation diagram for the HOMO energy levels of the molecule $TP-SiH_{2}-$ in the presence of electric field $E=\pm 0.001 au$. Due to the asymmetry of the molecule, the HOMO energy levels are asymmetrically displaced around the central phenyl ring. This is a consequence of the fact that $t_{1} \neq t_{2}$. For $E=0.001 au$ the HOMO is delocalized over the entire molecule indicating a reduction in the molecular admittance. In this case, the shift of the electron density is obviously less pronounced, since all the $\pi$- electrons are strongly involved in chemical bonds and transfer of an electron from one ring to the other becomes clearly much harder to achieve. On the other hand when $E=-0.001 au$ the HOMO energy levels are localized on the central and the right phenyl rings which indicates localization of the excess electron on the right phenyl ring. The $E_{v}$ for the molecule $TP-SiH_{2}-$ in the absence of electric field is estimated to be $3.25 eV$ which is lower than that of the previous molecule $TP-CH_{2}-$ ($3.59ev$).
A plot of energy gap $HLG$  versus electric field is shown in Fig.7. For the molecule $TP-CH_{2}-$, the plot is symmetric indicating the absence of any MRTD behaviour. On the other hand, for the second molecule $TP-SiH_{2}-$, the electron experiences a energy barrier on the positive electric field side. This basically indicate that on application of an electric field in the positive direction of the molecular axis, the electron experiences a higher barrier as compared to the case of an electric field in the negative direction of the molecular axis. This asymmetric response of the molecule to the external field is useful for possible molecular tunneling diode application. Replacing $Si$ with $Ge$ in the $TP-SiH_{2}-$ molecule does not significantly change the $HLG$ ($6.98 ev$) but on the other hand replacing $Si$ with $Sn$ significantly reduces the $HLG$ ($6.61 ev$). This is probably due to the metallic nature of $Sn$.

We also carried out DFT simulations on the $TP-SiH_{2}-$ molecule with $Cu-S$ metal atoms replacing the terminal $H$ atoms. The results indicate that the incorporation of the $Cu$ atoms decreases the HLG energy gap to $6.48eV$ due to the transfer of charge from $Cu$ atoms to the molecule.

\section{Conclusions}
Motivated by recent interest in the development of alternative to traditional electron devices, we have investigated a oligo-phenyl based double barrier molecular device. The double barriers have been created by using insulator spacers $-CH_{2}-$, $-SiH_{2}-$ , $GeH_{2}$ and $SnH_{2}$ inserted between two aromatic rings We have carried out analytical as well as numerical studies using HF and DFT methods. We found that when the insulator spacers are same the tunneling matrix elements of the barriers are identical and this leads identical current flowing in both directions of the molecule. The symmetry can be broken by making the two insulator spacers different. This leads to asymmetrical electron conduction along either direction of the molecule leading to the desired characteristics of a typical diode. The asymmetric conduction properties such as the one investigated in this study are relevant for the implementation of molecular tunneling diodes.

\section{Acknowledgements}

One of the authors Suman Dudeja thanks the University Grants Commission, New Delhi for the financial support for carrying out this work under grant no. $8-1(203)/(MRP/NRCB)$.

\end{document}